**Oxygen annealing induced changes in defects within (010) β-Ga$_2$O$_3$ epitaxial films measured using photoluminescence**

Rujun Sun[1], Yu Kee Ooi[1], Praneeth Ranga[1], Arkka Bhattacharyya[1], Sriram Krishnamoorthy[1], Michael A. Scarpulla [1,2]

[1.] *Electrical and Computer Engineering, University of Utah, Salt Lake City, UT, 84112, USA*

[2.] *Materials Science and Engineering, University of Utah, Salt Lake City, UT, 84112, USA*

**Abstract**

In this work, we use photoluminescence spectroscopy (PL) to monitor changes in the UV, UV', blue, and green emission bands from n-type (010) Ga$_2$O$_3$ films grown by metalorganic vapor phase epitaxy (MOVPE) induced by annealing at different temperatures under O$_2$ ambient. Annealing at successively higher temperatures decreases the overall PL yield and UV intensity at nearly the same rates, indicating the increase in formation of at least one non-radiative defect type. Simultaneously, the PL yield ratios of blue/UV and green/UV increase, suggesting that defects associated with these emissions increase in concentration with O$_2$ annealing. Utilizing the different absorption coefficients of 240 and 266 nm polarization-dependent excitation, we find an overall activation energy for the generation of non-radiative defects of 0.69 eV in the bulk but 1.55 eV near the surface. We also deduce activation energies for the green emission-related defects of 0.60 eV near the surface and 0.89-0.92 eV through the films, whereas the blue-related defects have activation energy in the range 0.43-0.62 eV for all depths. Lastly, we observe hillock surface morphologies and Cr diffusion from the substrate into the film for temperatures above 1050 °C. These observations are consistent with the formation and diffusion of V$_{Ga}$ and its complexes as a dominant process during O$_2$ annealing, but further work will be necessary to determine which defects and complexes provide radiative and non-radiative recombination channels and the detailed kinetic processes occurring at surfaces and in bulk amongst defect populations.

*Corresponding Authors Emails: rujun.sun@utah.edu and mike.scarpulla@utah.edu*

Ultra-wide bandgap β-$Ga_2O_3$ is promising in applications such as power electronic devices, solar blind UV photodetectors, solar cells and gas sensors[1]. Melt growth techniques such as edge-defined film fed growth[2] and Czochralski method[3] have been used to grow large-size single crystals, which sets β-$Ga_2O_3$ apart from other ultrawide gap materials like SiC, AlN, and diamond. The identification and manipulation of defects in $Ga_2O_3$ are important for its future applications. Thermal annealing under oxidizing atmospheres[2,4-6] (oxygen partial pressure $p_{O2}$ higher than the equilibrium vapor pressure of $O_2$ over $Ga_2O_3$[7]) is found to increase the resistivity of $Ga_2O_3$ crystals. The few hundred nm region near surface is more resistive than bulk after $O_2$ annealing[4,5] and its thickness can be suppressed by a capping $SiO_2$ layer[4]. The relation between resistivity and $p_{O2}$ also applies in single crystal growth[8] and thin film deposition[9]. Furthermore, under air annealing at 1450 °C for 20 h, O content of unintentionally-doped (UID) crystals increases while Ga content decreases as measured by EDS and XPS[6], indicating detectable mass change/loss[10].

For many decades, reduction in electron density upon annealing in oxidizing atmospheres is attributed to reduction in oxygen vacancy ($V_O$) concentration [8], which was assumed to be a shallow donor without direct evidence. However, density functional theory calculations using hybrid functions, which in recent years have significantly increased the accuracy of calculations of defect charge transition levels and other energies, indicate that $V_O$ should be a deep donor[11]. Thus, $V_O$ should not be the primary defect responsible for the resistivity increase under $O_2$ annealing. Note that Si and H impurities are leading candidates for the background n-type conductivity with theoretical[11] and experimental[12-14] evidence. Although direct experimental evidence of the microscopic changes in defect concentrations is sparse, positron annihilation spectroscopy (PAS) studies demonstrate that $V_{Ga}$ is efficiently formed during epitaxial growth of UID and Si-doped $Ga_2O_3$ films, and under $O_2$ annealing [15,16]. Electron paramagnetic resonance (EPR) spectra show $O_2$ annealing induces a IR1 center which is assigned to the double negative charge state of either isolated $V_{Ga}$ at the tetrahedral site or its relaxed configuration sometimes referred to as a $V_{Ga(I)}$-$Ga_{ib}$-$V_{Ga(I)}$ split vacancy complex[17]. Despite these basic insights, distinguishing main defects accounting for resistivity increases and fully investigating the formation, diffusion, and complexing mechanisms involved remain to be completed.

Photoluminescence spectroscopy is widely used to characterize defects. Four characteristic PL bands are observed for a majority of $\beta$-$Ga_2O_3$ samples upon super-gap excitation, namely, UV (centered at 3.4 eV), UV' (centered 3.0-3.2 eV), blue (centered 2.7-2.8 eV) and green (centered near 2.4 eV)[18]. Recent computations have shown that many different point defects and complexes offer internal transitions that should fall into the observed green-blue-UV' bandwidths and be vibronically broadened[19]. This may ultimately prevent unambiguous attribution of any of these peaks to one particular defect or complex. In addition, both wide and sharp red luminescence features are observed in some $Ga_2O_3$ crystals. The very sharp R1 and R2 lines very similar to ruby emissions as well as a broadened structure are caused by Cr[20,21] (not Fe) while some evidence exist for some different broad red features being associated with defects like N[22] and $V_{Ga}$-H acceptor[23].

The UV band's behavior is consistent with recombination of conduction electrons with self-trapped holes (STH)[24,25], while the UV', blue and green bands are generally ascribed to donor-acceptor pair recombination[26], although the involvement of STH especially in the UV' transition may also be possible. The ratios of blue and green (3.0 eV and 2.5 eV) peak yields to the UV yield are enhanced greatly after neutron irradiation[27], suggesting they arise from native defects. Galazka et al. observed that the blue emission in Sn-doped $Ga_2O_3$ crystal is greatly enhanced after $O_2$ annealing[10], but offered no further explanations. The green band is observed to be enhanced in the presence of specific impurities such as Be, Li, Ge and Sn[28], and with higher $p_{O2}$ during crystal growth or after $O_2$ annealing in UID crystals[29].

In this work, we use photoluminescence spectroscopy from MOVPE-grown homoepitaxial (010) β-$Ga_2O_3$ films subjected to $O_2$ annealing to investigate the defect-related processes during annealing. The ratios of blue and green peaks to the UV systematically increase with annealing temperature suggesting increases in the associated defect concentrations. Relying on PL interpretation along with prior positron annihilation, EPR and computational results, we associate the increased blue emission ratio to increased concentrations of compensating $V_{Ga}$ and its acceptor-like complexes. We derive activation energies for the process(es) introducing $V_{Ga}$-related defects into the films, which may include surface and bulk formation as well as bulk diffusion energies. Lastly, we observed isolated hillock surface morphologies and Cr diffusion from the substrate into the depth sampled by PL at sufficiently high temperatures, again supporting the hypothesis that the formation of $V_{Ga}$ and related complexes is the major process during annealing in high $p_{O2}$ atmospheres.

About 800 nm-thick Si-doped β-$Ga_2O_3$ films were grown by MOVPE[30] (Agnitron Agilis) on (010)-oriented NCT Fe-doped $Ga_2O_3$ substrates using triethyl gallium (TEGa), diluted silane and oxygen precursors. Annealing was conducted at 1 atm in a quartz tube furnace in which the heating rate and $O_2$ flow rate were 10 °C/min and 20 sccm respectively. Temperatures of 850, 900, 950, 1000, 1050 and 1100 °C were selected with the holding time of 2 hours. In order to control for any sample-to-sample variations, the same sample was measured in PL at 25 °C and then repeatedly annealed at the next increased temperature. Hall effect measurements (Ecopia HMS 7000) were used to characterize electrical properties of as-grown film (then the contacts were etched off) and after annealing at 1100 °C to avoid any contamination or alteration of the sample arising from adding and removing contacts between the intermediate annealing steps. Photoluminescence measurements were performed using both 266 nm and 240 nm excitations which provide depth selectivity. The polarization direction of the incident pulsed laser was rotated by a half-wave plate. The laser power was actively monitored (~25 mW, 0.07 cm$^2$). Emission spectra were collected in reflection mode using an integration sphere with a fiber connected to an Avantes spectrometer. For the analysis of PL spectra, the UV, UV', green and blue bands were fit using fixed peak energy and full-width at half-maximum.

The Si-doped as-grown film exhibits a carrier density of $3\times10^{17}$ cm$^{-3}$ and a mobility of 90 cm$^2$/(V.s). After $O_2$ annealing at 1100 °C, the resistivity exceeds the measurement limit of $10^6$ $\Omega$.cm, agreeing with the reported resistivity increase after $O_2$ annealing. All samples exhibited UV, UV', blue and green emission (Fig.1). The shape and intensity of PL spectra exhibit similar features under 240 nm excitation for annealing at 850 °C and 900 °C. As the annealing temperature increases, the total emission yield decreases continuously which we take as an indication of increasing non-radiative processes. Thus, the formation of at least one non-radiative defect center, whether in the bulk or at the surface, is the first defect-related process induced by annealing in $O_2$.

Before analyzing the PL spectra, normalizing the emission spectra to the UV emission is justified below. This should remove the influence of the non-radiative channels on excess carrier populations and thus we can consider each radiative channel. The absence of band-edge emission[31,32] indicate neither free holes nor its diffusion in valence band. The ultrafast dynamics experiments of photo-generated carriers show several ps and hundreds of ps timescales for STH[31] and holes trapped by defect[32], respectively. According to the phenomenological rate equation of holes after photo-excitation, holes will be captured by O sites, acceptor 1 and acceptor 2, and non-radiative centers by ignoring hopping of the bounded holes among them. Thus, the relative emission intensity is proportional to the relative defect density responsible for it since capture rate constant of defects should not change. Note that the detrapping energy for electrons and holes for the blue emission are 0.04 and 0.42 eV, respectively[26]. Another work obtained a similar 18 meV for electrons detrapping in blue emission[33]. Thus, the involved donor should be shallow like $Si_{Ga}$ with ionization energy in tens of meV, rather than deep donor like $V_O$. We thus propose that the blue emission is related to transitions between shallow donors and deep acceptors, so does the green emission. Due to the constant concentration of O sites, we could normalize others emission to UV emission. We recognize that understanding all subtleties in the PL spectra will require further consideration of the charge states of defects which is out of the scope of this study.

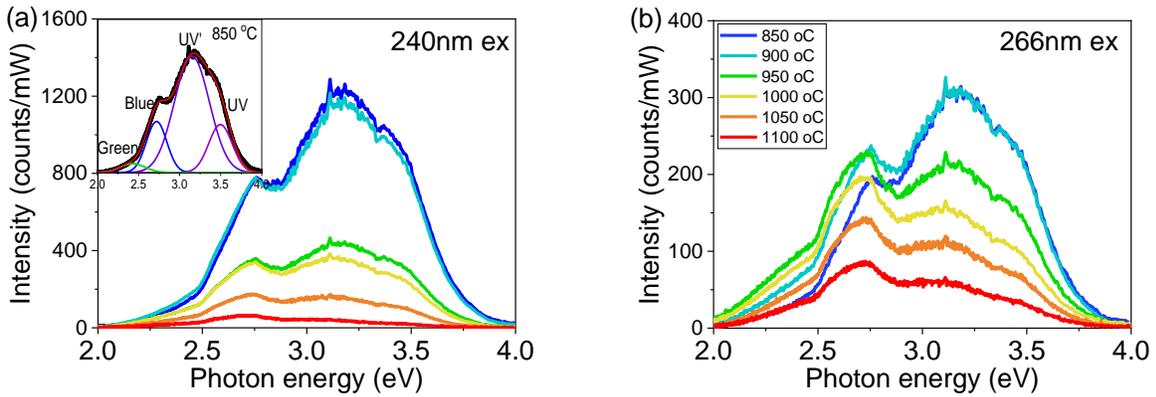

Figure 1. PL spectra of $Ga_2O_3$ film in E//c under (a) 240 nm and (b) 266 nm excitations.

The UV intensity changes in nearly the same way as the total PL yield (Fig.2a), indicating the dominant process to form self-trapped holes. The UV'/UV ratio remains nearly invariant as a function of annealing temperature (Fig.2b), which is consistent with assignment of the holes self-trapped at a different O site[18,19]. Transition from shallow donors to the self-trapped holes may also be involved in the UV' peak which would account for its lower energy and invariant UV'/UV ratio.

The decrease of carrier concentration (increased resistivity) upon annealing suggests a decrease of shallow donors or an increase in compensating acceptors. Si dopant will only diffuse instead of being deactivated under $O_2$ annealing above 1100 °C. A small amount of hydrogen, which could act as a donor as well as passivating acceptors, is possibly removed during oxygen annealing[34]. We have established via Hall, CV, and SIMS that Si is the dominant dopant in our films doped with Si, thus the hydrogen loss mechanism should not be dominant in our experiments[35]. Thus, the compensating acceptors are more likely to increase with annealing. The dependence of electron density on $O_2$ annealing arises from the acceptor density, and its PL intensity should go up with high acceptor concentrations. The relative intensities of the blue and green emissions do change as a function of annealing temperature. Specifically, $V_{Ga}$ and its complexes are most probable given the existing literature and our observations. In computations for defect formation in the bulk, $V_{Ga}$ exhibits the lowest formation energy amongst native acceptors under oxygen rich condition[36,37], especially for n-type doped films in which the Fermi energy approaches the conduction band. Combining with the facts of the reported PAS results[15,16] and the indicated intrinsic defect nature by neutron irradiation[27], ascribing the increased blue emission to increases in the concentrations of $V_{Ga}$ and related complexes is consistent with the mentioned findings.

Enhanced green peak might also come from intrinsic defects, as revealed by the growth of crystals in high $O_2$ partial pressure[29] and this work under $O_2$ annealing. Compared to blue emission energy, the lower emission energy indicates defects involved could be deeper acceptor (cannot rule out deeper donor). More work is needed to demonstrate the correlation between green emission and related defect, and its correlation to the electrical properties of $Ga_2O_3$.

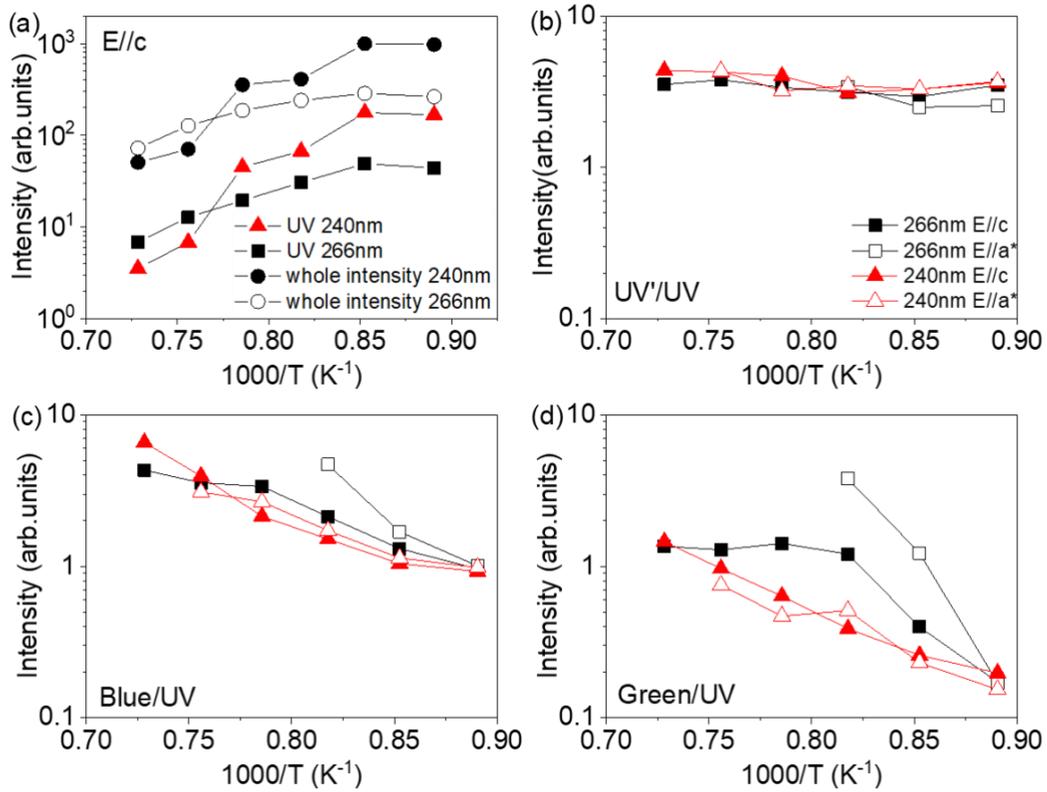

**Figure 2**. (a) overall integrated intensity along with UV intensity, ratios of (b) UV', (c) blue and (d) green to UV (integrated area) for annealing at different temperatures.

In the absence of mass transport to or from the surface, smoothening to reduce surface area rather than increasing it would be expected. In Fig.3, changes in the surface morphology from linear features to isolated hillock features after annealing at 1100 °C do hint a significant mass transport processes involving the surface. No etching or decomposition and desorption of suboxides is expected at these temperatures, pressures, and $p_{O2}$ according to thermodynamics[7].

The 1 atm of pure $O_2$ overpressure implies ~$10^9$ $O_2$ atoms/s would bombard each surface site. Such oxygen flux would be sufficient to grow a similar number of monolayers of $Ga_2O_3$ per second were sufficient Ga supplied. Given these considerations, we propose that this change in morphology results from gallium atoms diffusing to the surface and reacting with little adatom diffusion with the copious impinging $O_2$ to form new unit cells of $Ga_2O_3$ in hillocks as opposed to smooth layer by layer growth. If the outdiffusion mechanism for Ga is via substitutional Ga or ($2V_{Ga} + Ga_i$) complexes[38], it could equivalently be viewed as the formation of $V_{Ga}$ or the complex at the surface having one activation energy followed by its in-diffusion under another activation energy.

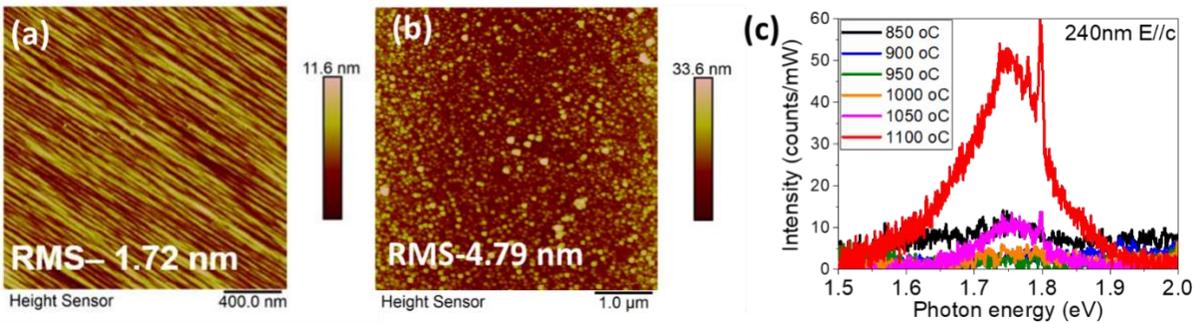

Figure 3 - atomic force microscope images of films (a) as-grown (similar growth condition) and (b) after annealing at 1100 °C. (c) PL spectra of red emissions from $Ga_2O_3$ epitaxial films annealed at different temperatures

Interestingly, the red luminescence originating from $Cr^{21}$ is observed near top surface after annealing at 1050 °C and 1100 °C (Fig. 3c), but not at lower temperatures. This correlates with the diffusion of Cr from the Fe-doped substrate into the film. Note that all of the ion-implanted $Si^{13,39,40}$, $Sn^{39,41}$, $Ge^{41}$ are observed to diffuse near 1100 °C in n-type $Ga_2O_3$, in spite of these donors having different radii, signifying that their diffusion is limited by the same activation energy ($E_a$).

This process is consistent the formation of $V_{Ga}$ and diffusion through $V_{Ga}$ since all of these dopants are expected to substitutionally sit on Ga sites including Cr[42]. On the other hand, the diffusions of Si, Sn, Ge during $O_2$ annealing are faster than that in $N_2$ annealing[40]. This can be explained by the higher concentration of $V_{Ga}$ in O-rich condition compared to O-poor one, consistent with our results and theoretical indication[37]. The ion implanted Si, Sn and Ge are concentrated near surface (<20 nm) upon $O_2$ annealing[39-41], indicating that the formation energy of $V_{Ga}$ at surface is smaller than that in bulk. The situation could be that the $V_{Ga}$ forms at surface and diffuses into bulk, while the Cr impurities diffuse into films by exchanging position with $V_{Ga}$, thus moving towards surface.

Table 1 shows the extracted activation energy ($E_a$) for the UV intensity and the blue/UV and green/UV intensity ratios vs. $1/T_{anneal}$. The $E_a$ of UV absolute intensity is 1.55 eV near surface and 0.64 eV in deep film, which represents the quenching of UV peaks by increased non-radiative recombinations accompanying higher-temperature annealing. The $E_a$ extracted from green emission are higher than that of blue emission, suggesting different defect origins or mechanisms between them. Considering the as-mentioned experimental phenomena, we try to find evidence that the surface formation energy is relatively small and the bulk diffusion process follows a higher barrier value, in the depth (b axis) direction. The blue peak seems to exhibit only slight change of $E_a$ while the green has a stronger surface vs bulk contrast. Note that the value $E_a$ of 266 nm E//a* is omitted for this analysis since it is in low intensity thus large error and also contains signal from substrate. Further work will be necessary to determine more accurate activation energies for their formation, reactions and transport at the surface and within the bulk.

Table 1 The activation energies fitted by $N_0 \exp(-E_a/k_B T)$. Penetration depth is calculated using the absorption coefficients of (010) crystal[43].

| Excitation | Penetration depth (nm) | $E_a$ (eV) | | |
|---|---|---|---|---|
| | | UV | blue/UV | green/UV |
| 240nm E//c | 30 | 1.55±0.08 | 0.62±0.06 | 0.60±0.03 |
| 240nm E//a* | 50 | - | 0.46±0.03 | 0.89±0.15 |
| 266nm E//c | 500 | 0.69±0.06 | 0.43±0.02 | 0.92±0.06 |

Since there is no reported value on $V_{Ga}$ formation and diffusion energy, we compare it with other processes. Mg starts diffusing already through interstitial Ga[44] at 800 °C[45], indicating a smaller diffusion barrier. The predicted $E_a$ for Mg diffusion (slightly > 2 eV) [44] is actually higher than our extracted $E_a$ related to $V_{Ga}$. This is likely due to an offset between computational and experimental values. Our extracted values are also smaller than the 4.1 eV extracted using the thickness of semi-insulating layer as a function of temperature and time using equivalent circuit model of CV measurements[4]. It is reasonable that the extracted $E_a$ differs due to different physical processes involved. In addition, an activation energy of 0.50~0.65 eV is extracted by thermally stimulated luminescence spectroscopy and identified as the deep level of a compensating acceptor in heavily Si-doped but semi-insulating $Ga_2O_3$ films grown by MOVPE[46]. This is somewhat consistent with our findings.

In conclusion, we investigated the photoluminescence of Si-doped (010) β-$Ga_2O_3$ MOVPE films as they become semi-insulating under $O_2$ annealing. The overall PL yield and the UV yield decrease in the same ways upon annealing. The blue and green to UV intensity ratios exponentially increase with annealing temperature while the UV'/UV ratio remains invariant. We extract an activation energy of 0.43-0.62 eV for blue emission for all depths in the film suggesting a bulk process, while the green emission data hints at differentiation between surface and bulk. Cr impurities are observed to diffuse into the epitaxial film at 1050~1100 $^{o}$C, possibly by substitutional diffusion mediated by $V_{Ga}$. Our findings, including the morphology change of the surface seen in AFM, are all consistent with the formation of $V_{Ga}$ and its complexes as a major defect process during $O_2$ annealing. Note that our results cannot prove that they alone are the cause of the decreased electron concentration, lower overall PL yields, and changes in blue and green emissions.

This material is based upon work supported by the Air Force Office of Scientific Research under award number FA9550-18-1-0507 (Program Manager: Dr. Ali Sayir). Any opinions, finding, and conclusions or recommendations expressed in this material are those of the author(s) and do not necessarily reflect the views of the United States Air Force. The authors thank Prof. Steve Blair at the University of Utah for providing polarized photoluminescence measurement facilities.

The data that support the findings of this study are available from the corresponding author upon reasonable request.

**References**

[1] S. J. Pearton, Jiancheng Yang, Patrick H. Cary, F. Ren, Jihyun Kim, Marko J. Tadjer, and Michael A. Mastro, Applied Physics Reviews **5** (1), 011301 (2018).
[2] Akito Kuramata, Kimiyoshi Koshi, Shinya Watanabe, Yu Yamaoka, Takekazu Masui, and Shigenobu Yamakoshi, Jpn. J. Appl. Phys. **55** (12), 1202A2 (2016).
[3] Zbigniew Galazka, Reinhard Uecker, Detlef Klimm, Klaus Irmscher, Martin Naumann, Mike Pietsch, Albert Kwasniewski, Rainer Bertram, Steffen Ganschow, and Matthias Bickermann, ECS J. Solid State Sci. Technol. **6** (2), Q3007 (2016).

4   Takayoshi Oshima, Kenichi Kaminaga, Akira Mukai, Kohei Sasaki, Takekazu Masui, Akito Kuramata, Shigenobu Yamakoshi, Shizuo Fujita, and Akira Ohtomo,  Jpn. J. Appl. Phys. **52** (5R), 051101 (2013).
5   Zbigniew Galazka, Klaus Irmscher, Reinhard Uecker, Rainer Bertram, Mike Pietsch, Albert Kwasniewski, Martin Naumann, Tobias Schulz, Robert Schewski, Detlef Klimm, and Matthias Bickermann,  J. Cryst. Growth **404**, 184 (2014).
6   Huiyuan Cui, Qinglin Sai, Hongji Qi, Jingtai Zhao, Jiliang Si, and Mingyan Pan,  J Mater Sci **54** (19), 12643 (2019).
7   D. Klimm, S. Ganschow, D. Schulz, R. Bertram, R. Uecker, P. Reiche, and R. Fornari,  J. Cryst. Growth **311** (3), 534 (2009).
8   Naoyuki Ueda, Hideo Hosono, Ryuta Waseda, and Hiroshi Kawazoe,  Appl. Phys. Lett. **70** (26), 3561 (1997).
9   Masahiro Orita, Hiromichi Ohta, Masahiro Hirano, and Hideo Hosono,  Appl. Phys. Lett. **77** (25), 4166 (2000).
10  Z. Galazka,  Semicond. Sci. Technol. **33** (11) (2018).
11  J. B. Varley, J. R. Weber, A. Janotti, and C. G. Van de Walle,  Appl. Phys. Lett. **97** (14), 142106 (2010).
12  Encarnación G. Víllora, Kiyoshi Shimamura, Yukio Yoshikawa, Takekazu Ujiie, and Kazuo Aoki,  Appl. Phys. Lett. **92** (20), 202120 (2008).
13  Kohei Sasaki, Masataka Higashiwaki, Akito Kuramata, Takekazu Masui, and Shigenobu Yamakoshi,  Appl. Phys. Express **6** (8), 086502 (2013).
14  P. D. C. King, I. McKenzie, and T. D. Veal,  Appl. Phys. Lett. **96** (6), 062110 (2010).
15  E. Korhonen, F. Tuomisto, D. Gogova, G. Wagner, M. Baldini, Z. Galazka, R. Schewski, and M. Albrecht,  Appl. Phys. Lett. **106** (24), 242103 (2015).
16  J. Tadjer Marko, Freitas Jaime, Culbertson James, Weber Marc, R. Glaser Evan, Mock Alyssa, Mahadik Nadeemullah, J. Schmieder Kenneth, Jackson Eric, Gallagher James, Feigelson Boris, and Kuramata Akito,  J. Phys. D: Appl. Phys. (2020).
17  Nguyen Tien Son, Quoc Duy Ho, Ken Goto, Hiroshi Abe, Takeshi Ohshima, Bo Monemar, Yoshinao Kumagai, Thomas Frauenheim, and Peter Deák,  Appl. Phys. Lett. **117** (3), 032101 (2020).
18  Yunshan Wang, Peter T. Dickens, Joel B. Varley, Xiaojuan Ni, Emmanuel Lotubai, Samuel Sprawls, Feng Liu, Vincenzo Lordi, Sriram Krishnamoorthy, Steve Blair, Kelvin G. Lynn, Michael Scarpulla, and Berardi Sensale-Rodriguez,  Scientific Reports **8** (1), 18075 (2018).
19  Y. K. Frodason, K. M. Johansen, L. Vines, and J. B. Varley,  J. Appl. Phys. **127** (7), 075701 (2020).
20  Yoshinori Tokida and Sadao Adachi,  J. Appl. Phys. **112** (6), 063522 (2012).
21  Rujun Sun, Yu Kee Ooi, Peter T. Dickens, Kelvin G. Lynn, and Michael A. Scarpulla,  Appl. Phys. Lett. **117** (5), 052101 (2020).
22  Y. P. Song, H. Z. Zhang, C. Lin, Y. W. Zhu, G. H. Li, F. H. Yang, and D. P. Yu,  Phys. Rev. B **69** (7), 075304 (2004).
23  Thanh Tung Huynh, Ekaterine Chikoidze, Curtis P Irvine, Muhammad Zakria, Yves Dumont, Ferechteh H Teherani, Eric V Sandana, Philippe Bove, David J Rogers, and Matthew R Phillips,  arXiv preprint arXiv:2005.13743 (2020).
24  G. Blasse and A. Bril,  J. Phys. Chem. Solids **31** (4), 707 (1970).
25  T. Harwig, F. Kellendonk, and S. Slappendel,  J. Phys. Chem. Solids **39** (6), 675 (1978).
26  Laurent Binet and Didier Gourier,  J. Phys. Chem. Solids **59** (8), 1241 (1998).
27  H. Gao, S. Muralidharan, N. Pronin, M. R. Karim, S. M. White, T. Asel, G. Foster, S. Krishnamoorthy, S. Rajan, L. R. Cao, M. Higashiwaki, H. Von Wenckstern, M. Grundmann, H. Zhao, D. C. Look, and L. J. Brillson,  Appl. Phys. Lett. **112** (24) (2018).
28  T. Harwig and F. Kellendonk,  J. Solid State Chem. **24** (3), 255 (1978).

29 Encarnación G. Víllora, Mitsuo Yamaga, Takafumi Inoue, Satoshi Yabasi, Yuki Masui, Takasi Sugawara, and Tsuguo Fukuda, Jpn. J. Appl. Phys. **41** (Part 2, No. 6A), L622 (2002).

30 Praneeth Ranga, Arkka Bhattacharyya, Ashwin Rishinaramangalam, Yu Kee Ooi, Michael A. Scarpulla, Daniel Feezell, and Sriram Krishnamoorthy, Appl. Phys. Express **13** (4), 045501 (2020).

31 Saulius Marcinkevičius and James S. Speck, Appl. Phys. Lett. **116** (13), 132101 (2020).

32 Okan Koksal, Nicholas Tanen, Debdeep Jena, Huili Xing, and Farhan Rana, Appl. Phys. Lett. **113** (25), 252102 (2018).

33 Katerina M. Othonos, Matthew Zervos, Constantinos Christofides, and Andreas Othonos, Nanoscale Res. Lett. **10** (1), 304 (2015).

34 Zbigniew Galazka, Klaus Irmscher, Robert Schewski, Isabelle M. Hanke, Mike Pietsch, Steffen Ganschow, Detlef Klimm, Andrea Dittmar, Andreas Fiedler, Thomas Schroeder, and Matthias Bickermann, J. Cryst. Growth **529**, 125297 (2020).

35 Arkka Bhattacharyya, Praneeth Ranga, Saurav Roy, Jonathan Ogle, Luisa Whittaker-Brooks, and Sriram Krishnamoorthy, arXiv preprint arXiv:2008.00303 (2020).

36 T. Zacherle, P. C. Schmidt, and M. Martin, Phys. Rev. B **87** (23) (2013).

37 M. E. Ingebrigtsen, A. Yu Kuznetsov, B. G. Svensson, G. Alfieri, A. Mihaila, U. Badstübner, A. Perron, L. Vines, and J. B. Varley, APL Mater. **7** (2), 022510 (2019).

38 Jared M. Johnson, Zhen Chen, Joel B. Varley, Christine M. Jackson, Esmat Farzana, Zeng Zhang, Aaron R. Arehart, Hsien-Lien Huang, Arda Genc, Steven A. Ringel, Chris G. Van de Walle, David A. Muller, and Jinwoo Hwang, Physical Review X **9** (4), 041027 (2019).

39 Marko J. Tadjer, Chaker Fares, Nadeemullah A. Mahadik, Jaime A. Freitas, David Smith, Ribhu Sharma, Mark E. Law, Fan Ren, S. J. Pearton, and A. Kuramata, ECS J. Solid State Sci. Technol. **8** (7), Q3133 (2019).

40 Ribhu Sharma, Mark E. Law, Chaker Fares, Marko Tadjer, Fan Ren, A. Kuramata, and S. J. Pearton, AIP Advances **9** (8), 085111 (2019).

41 Ribhu Sharma, Mark E. Law, Minghan Xian, Marko Tadjer, Elaf A. Anber, Daniel Foley, Andrew C. Lang, James L. Hart, James Nathaniel, Mitra L. Taheri, Fan Ren, S. J. Pearton, and A. Kuramata, Journal of Vacuum Science & Technology B **37** (5), 051204 (2019).

42 Varley J.B.

43 Takeyoshi Onuma, Shingo Saito, Kohei Sasaki, Tatekazu Masui, Tomohiro Yamaguchi, Tohru Honda, Akito Kuramata, and Masataka Higashiwaki, Jpn. J. Appl. Phys. **55** (12), 1202B2 (2016).

44 Hartwin Peelaers, John L. Lyons, Joel B. Varley, and Chris G. Van de Walle, APL Mater. **7** (2), 022519 (2019).

45 Man Hoi Wong, Chia-Hung Lin, Akito Kuramata, Shigenobu Yamakoshi, Hisashi Murakami, Yoshinao Kumagai, and Masataka Higashiwaki, Appl. Phys. Lett. **113** (10), 102103 (2018).

46 Md Minhazul Islam, Naresh Adhikari, Armando Hernandez, Adam Janover, Steven Novak, Sahil Agarwal, Charles L. Codding, Michael Snure, Mengbing Huang, and Farida A. Selim, J. Appl. Phys. **127** (14), 145701 (2020).